\documentclass[referee]{raa}            % referee version: for submission

\usepackage{graphicx,times}             %for PS/EPS graphics inclusion, new
\usepackage{CJKutf8}
\usepackage{siunitx}
\usepackage{natbib}
\usepackage{amssymb,amsmath}
\bibpunct{(}{)}{;}{a}{}{,}

\usepackage{multirow}
\definecolor{first}{RGB}{192, 226, 202}
\definecolor{second}{RGB}{227, 237, 186}
\definecolor{third}{RGB}{255, 247, 188}

\usepackage[a4paper]{geometry}
\usepackage[pagebackref=true]{hyperref}

\begin{document}
\begin{CJK*}{UTF8}{gbsn}
  \title{Enhancing astrometric registration of Chinese historical Astronomical Digital Plates with deep learning}
%   \subtitle{I. Place Your Subtitle Here}

   \volnopage{Vol.0 (20xx) No.0, 000--000}%%preserved for Editor. DOn't remove!
   \setcounter{page}{1}          %%starting page, preserved for Editor. DOn't remove!

   \author{Quanfeng Xu (徐权峰) %% Put your Chinese name in "( )" if you like. Note to open line 11 "\usepackage[UTF8]{ctex}"
      \inst{1,2}
   \and Zhengjun Shang (商正君)
      \inst{1}
   \and Shiyin Shen (沈世银)
      \inst{1,3}
   \and Yong Yu (于涌)
      \inst{1,2}
   \and Meiting Yang (杨美婷)
      \inst{1}
   \and Hao Luo (罗浩)
      \inst{1}
   \and Zhenghong Tang (唐正宏)
      \inst{1,2}
   \and Jing Yang (杨静)
      \inst{1}
   \and Jianhai Zhao (赵建海)
      \inst{1}
   }

   \institute{Shanghai Astronomical Observatory, Chinese Academy of Sciences, 80 Nandan Rd., Shanghai 200030, China; {\it ssy@shao.ac.cn, \it yuy@shao.ac.cn, \it ymt@shao.ac.cn}\\
        \and
             School of Astronomy and Space Science, University of Chinese Academy of Sciences, 1 East Yanqi Lake Rd., Beijing 100049, P.R. China;\\
        \and
             Key Lab for Astrophysics, Shanghai, 200234, China\\
\vs\no
   {\small Received 20xx month day; accepted 20xx month day}}

\abstract{China has systematically collected nighttime astronomical plates since 1900, creating a large historical dataset that has been digitized with optical scanners. For astrometric registration of these digitized plates, sources were first extracted using \texttt{SExtractor}, and then matched astrometrically with \texttt{Astrometry.net} and the \texttt{Gaia} catalog. However, suboptimal early storage conditions and subsequent environmental deterioration have impeded accurate source matching, resulting in processing failures for several thousand digitized plates. In this work, we introduce a Transformer-based classification model that takes cutouts of \texttt{SExtractor}-detected sources as input and leverages multi-scale feature fusion to identify trustworthy stellar sources on the plates. Trained on plates with successful astrometric calibration, our AI-based classifier was then applied to \texttt{SExtractor} detected sources of \num{1883} digitized plates, enabling us to complete the astrometric registration for \num{1353} of them. This AI-augmented pipeline streamlines the processing of historical plate archives and enhances their scientific value for long-term time-domain astronomical studies.
}

   \authorrunning{Xu et al.}            %author_head in even pages
   \titlerunning{Astronomical Plates Calibration}  % title_head in odd pages

   \maketitle
%% The author head (on even pages) and the title head (on odd pages) will be
%% automatically extracted from \author{} and \title{}. Whenever the title is too long,
%% you will be asked to supply a shorter one by inserting either \authorrunning{} or
%% \titlerunning{} before \maketitle. Anyway, you can specify your own heads.
%%
%%
%% Note: In the following text body of your manuscript, please note several differences from
%%       other major journals:
%% (1) \subsection{Please Capitalize the First Letter of Each Notional Word in Subsection Title}
%% (2) Please Capitalize the First Letter of Each Notional Word in all tables' captions

%
%________________________________________________ sections below
%
\section{Introduction}

Astronomical plates represent one of the most important observational assets of twentieth-century astronomy, providing long-term records of the sky that modern surveys cannot replicate. With the advent of high-performance hardware scanners, these fragile photographic materials can now be digitized with high geometric and photometric precision. The Harvard DASCH project \cite[][]{Grindlay_Tang_Los_Servillat_2011}, for example, employed a custom-built ultra-precise scanning system \cite[][]{simcoe2006ultrahigh} to digitize about \num{450000} plates obtained from 1890 to 1990. %#有没有欧洲的相关文章，比如supercosmos#
In parallel, the APPLAUSE project \cite[][]{2024A&A...687A.165E} (Leibniz Institute for Astrophysics Potsdam) has released high-precision scans and catalogs for over \num{70000} historical plates (1880–1999) obtained with commercial flatbed scanners, providing 2 billion calibrated photometric measurements. Similarly, the SuperCOSMOS Sky Survey \cite[][]{10.1111/j.1365-2966.2001.04660.x, 10.1111/j.1365-2966.2001.04661.x} has fully digitized 1.3 million plates from the UK and ESO Schmidt telescopes (1950s–1990s), delivering the reference sub-arcsecond, multi-band catalog for the southern sky.

Inspired by such efforts, China has launched its own digitization programs, including the high-performance digitizer developed at the Shanghai Astronomical Observatory \cite[][]{Yu_2017}, which achieves submicron-level positional precision through repeated scans and rigorous geometric and photometric corrections. Following digitization, each plate needs to be astrometrically registered for astronomical study, which requires accurately determining its pointing, scale, and position angle by cross-matching detected stellar sources with modern reference catalogs such as \texttt{Gaia}. In \cite{Shang_2024}, for all \num{29314} Chinese historical nighttime astronomical plates that have been digitized, \num{13512} out of \num{18226} single-exposure plates were successfully registered for their astrometry by matching the stellar sources detected on the digitized images with the \texttt{Gaia DR2} catalog. Additionally, the photometric calibration for these plates has also been completed by \cite{Ma_2025}.

However, many historical plates suffer from scratches, emulsion degradation, uneven backgrounds, or exposure variations, which complicate the reliable detection of stellar sources and hinder follow-up astrometric registration. To systematically address and quantify the impact of the degradations, a plate quality grading scheme has been established by \cite{Shang_2024}(see Section~\ref{sec:uncaldata} for detail).% By benchmarking our classification model across different quality grades, we aim to demonstrate its robustness in handling various degrees of emulsion damage and mold.}
This difficulty highlights a key issue in the astrometric registration workflow: the need to distinguish real stellar sources from artifacts. In \cite{Shang_2024}, they employed a support vector machine classifier \cite[SVM,][]{hearst1998support} to further classify the detected sources into stellar and non-stellar objects using the photometric parameters output from \texttt{SExtractor}. Using the stellar sources purified by this SVM classifier, \citet{Shang_2024} achieved a successful astrometric calibration for additional \num{2184} plates.

%Despite these challenges, most plates still contain a sufficient number of well-preserved stellar images with stable point-spread characteristics. These high-quality stars provide robust anchors for astrometric calibration, enabling accurate recovery of plate calibration even when large portions of the field are degraded. Severely degraded regions remain challenging for traditional pipelines and often lead to unreliable or failed solutions.

 %This makes accurate star or non-star classification a critical prerequisite for successful astrometric reduction. 
%This issue is exemplified by the results of \cite{Shang_2024}, despite obtaining successful matches for 

%To address the remaining \num{4714} unmatched images, t Nevertheless, \num{2530} plates remain unresolved, as plate contamination significantly complicates source identification.

Despite these efforts, there are still \num{2530} single-exposure Chinese historical nighttime astronomical plates that remain to be astrometrically calibrated. These astrometrically uncalibrated plates prompted us to adopt a new classification method to classify the stellar objects more effectively. In fact, modern computer vision methods based on deep learning have demonstrated superior classification performance compared to traditional machine learning approaches, such as SVMs based on engineered features. For example, \cite{slater2020morphological} employed a CNN-based method to effectively classify galaxies and stars in crowded stellar fields, and \cite{magnier2020pan} applied a CNN approach to transient classification in wide-field surveys. Additionally, \cite{walmsley2020galaxy} utilized Galaxy Zoo data \cite[][]{fortson2012galaxy} to perform a complex morphological classification of galaxies \cite[see also][]{Xu2023from,10.1093/mnras/staf025}.
%事实上，基于深度学习的现代计算机视觉方法已经表现出超越传统的机器学习方法（如基于特征工程的SVM）的分类能力。
These successful applications prompted us to develop a deep-learning-based classifier for stellar and non-stellar objects in the scanned images of Chinese historical nighttime astronomical plates that remain uncalibrated.

In this study, we use the transformer-based classification framework to accomplish this task. The primary difficulty lies in the fact that stellar sources exhibit stable photometric profiles determined by the telescope’s optics and exposure parameters, whereas artifacts produced by scratches, dust, or emulsion defects can differ from plate to plate. To address this complex binary classification problem, we first train a baseline model using data from plates that have already undergone astrometric registration. Building on this, for images that the baseline model misclassifies, we will construct customized training sets with similar physical characteristics to further fine-tune and optimize the model, thereby completing the final classification task.

The structure of this paper is as follows: Section~\ref{sec:method_data} describes the data and methodology, including the astrometric registration workflow, the Swin Transformer classifier, and the construction of training and validation datasets. Section~\ref{sec:results} presents the training and evaluation of the base model, its application to uncalibrated plates, and the fine-tuning strategy for challenging cases. Finally, section~\ref{sec:discussion} specifically discusses the cases that still cannot be calibrated, and section~\ref{sec:summary} summarizes the conclusions.

\section{Method and Data}\label{sec:method_data}

\begin{figure*}
    \includegraphics[width=\linewidth]{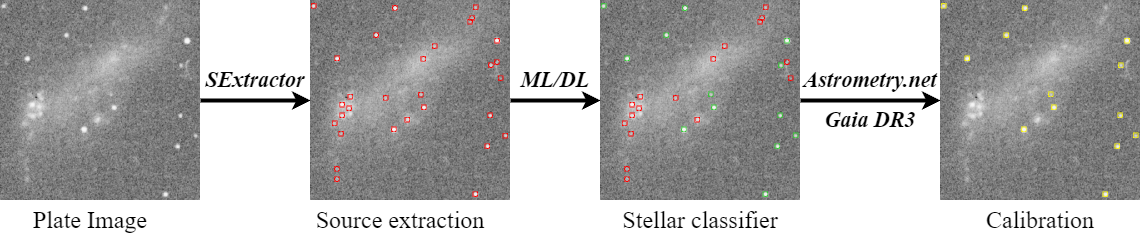}
    \caption{
    An overview of the astrometric registration workflow for Chinese historical nighttime astronomical plates (Example for plate BJ8703SD2899001).
    }
    \label{fig:traditional_method}
\end{figure*}

\subsection{Astrometric registration workflow}\label{sec:calibration_workflow}

The astrometric registration of the digitized plates consists of three main steps, as shown in Figure~\ref{fig:traditional_method}: source extraction, stellar source classification, and astrometric matching.% 找源，分类，匹配（分两步）
\begin{itemize}

\item \textit{Source extraction}: 
In this stage, the objective is to identify all statistically significant intensity enhancements on the digitized plate and characterize them as candidate astronomical sources. The process produces an initial catalog containing positional and photometric measurements that serve as inputs for downstream analysis.
In this study, \texttt{SExtractor} is used to detect raw astronomical sources on the digitized plate image \cite{refId0}. As an example, we show the source extraction result for a plate in the second panel of Fig.~\ref{fig:traditional_method}, where the detected sources are marked with \fcolorbox{red}{white}{red boxes}.

\item \textit{Stellar Source Classification}: 
This step identifies the genuine stellar sources from the non-stellar objects and spurious detections. 
This stellar source classification process could be accomplished by either using the photometric parameters output from \texttt{SExtractor} \cite[e.g., SVM in][]{Shang_2024} or the cutout images, as will be done in this study.
As an example, we show the stellar source classification for the plate in the third panel of Fig.~\ref{fig:traditional_method}, where the stellar sources identified by \cite{Shang_2024} are marked with \fcolorbox{green}{white}{green boxes}.

%The vetted stellar sources provide stable positional anchors, enabling robust astrometric calibration even on plates affected by significant photographic degradation.
%A machine learning classifier is employed to distinguish stars based on their characteristic PSF and photometric properties, with stellar sources highlighted in \fcolorbox{green}{white}{green boxes} and non-stellar sources in \fcolorbox{red}{white}{red boxes}.

\item \textit{Astrometric matching}: 
This step matches the stellar sources obtained previously with a modern astrometric catalog (e.g., \texttt{Gaia}) to derive the astrometric coordinates of the entire plate. The process consists of two iterative stages: coarse matching is performed using tools such as \texttt{Astrometry.net} \citep{Lang_2010}, providing an initial World Coordinate System (WCS) estimate; then, building on this coarse match, epoch information is incorporated to refine the astrometric calibration of the plate and the precise positions of stars. The final calibrated solution achieves alignment with high-precision catalogs, as illustrated in the rightmost panel of Fig.~\ref{fig:traditional_method}, where stellar sources fully aligned with \texttt{Gaia DR3} astrometric positions are marked with \fcolorbox{yellow}{white}{yellow boxes}.
% 这一步是将上一步得到的恒星源和现代天体测量星表（如Gaia）进行位置匹配来获得整个底片的天体测量坐标。具体来说，该过程分为两个可迭代步骤：首先利用A进行粗匹配；然后在粗匹配基础上结合历元信息获取底片（天体测量）及恒星目标的精确天体测量位置。例如，最右的图中的我们给出了最终和Gaia DR3给出的天体测量位置完全相匹配的恒星目标（box）

\end{itemize}

\begin{figure*}
    \centering
    \includegraphics[width=0.95\linewidth]{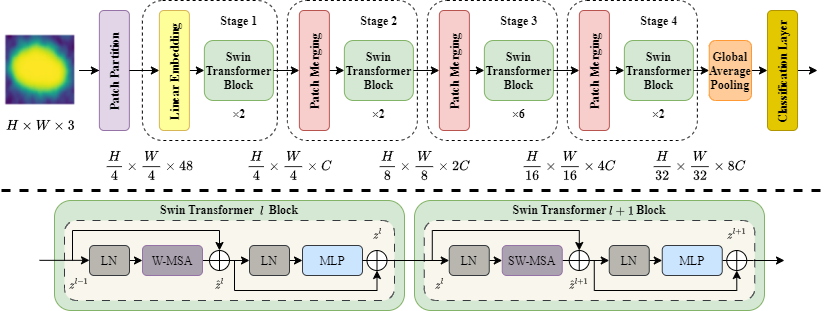}
    \caption{
    An overview of Swin Transformer methods includes the structural details.
    }
    \label{fig:swin}
\end{figure*}

Among these steps, the stellar classification stage provides the crucial link between the raw scan image and astrometric registration, ensuring that trustworthy stellar sources offer precise positional anchors.

\subsection{Swin Transformer Classifier}\label{sec:classifier}

% However, plates with severe degradation (e.g., scratches, mold, or emulsion damage) frequently fail to solve with \texttt{Astrometry.net} because a large fraction of \texttt{SExtractor} detected sources are artifacts rather than genuine stars. To address this, we insert an AI-based source-filtering step prior to astrometric calibration (ML/DL in Figure~\ref{fig:traditional_method}).% 与上面重复。
% ##这儿缺一段逻辑上的过渡，就是说为了解决假源问题，我们在source identification这一步里面增加一步AI分类器，具体结构如下。
% To improve alignment, we incorporate an AI-based classifier into the source identification step to distinguish genuine sources from defects, as illustrated by the red line in Figure~\ref{fig:traditional_method}.

We adopt the Swin Transformer backbone \cite{liu2021swin} as the core classifier (see the architecture in Figure~\ref{fig:swin}). The model uses a hierarchical vision transformer \cite{dosovitskiy2021an} with shifted windows: The architecture begins by partitioning the input patch ($64\times64$ pixels) into $4\times4$ non-overlapping patches, which are linearly embedded ($C=48$). Secondly, the Swin Transformer Blocks within each stage perform efficient self-attention computation, first within regular non-overlapping window-based multi-head self-attention (W-MSA) and then within shifted-window (SW-MSA) in the next block, enabling cross-window information exchange. Finally, a global average pooling layer followed by a classifier produces the binary classification and optimizes the model with a weighted cross-entropy loss function.

% ##这个Swin Transformer Classifer的输入是X*X大小的图片cutout，输出是一个简单的二分类。我们用？计算模型的损失。

%In this pipeline, each candidate source is cropped into a centered patch and fed into the frozen Swin Transformer model for inference. This simple filtering step advances the robust astrometric calibration without requiring any manual intervention.

\subsection{Training Sample from Calibrated Plates}\label{sec:traindata}

\begin{table}
\centering
\caption{Status and processing of Chinese historical nighttime astronomical plates by telescopes}
% \begin{tabular}{|c|c|c|c|c|c|}
\setlength{\tabcolsep}{1.75mm}{
\begin{tabular}{cccccc}
\hline
\multirow{2}*{\textbf{Observatory}} & \multirow{2}*{\textbf{Telescope}} & \textbf{Calibrated} & \textbf{To be}& \textbf{Base}&\textbf{Fine-tune}\\ 
 &  & \textbf{by \citeauthor{Shang_2024}}%\textcolor{blue}{Shang}
 & \textbf{Calibrated}& \textbf{Model}&\textbf{Model} \\ 
\hline
\multirow{3}*{SHAO} & 40cm double-tube refractor & 1328 & 470& 283& 20 \\
  & 1.56m reflector telescope & 167 & 10 & 3 & 1 \\
  & No record & 44 & 6& - & - \\
\hline
\multirow{3}*{QDO} & 32cm refractor telescope & 81 & 75 & 59 & 7\\
  & 15cm refractor telescope & - & 11 & 5 & - \\
  & No record & 60 & 31& 21& - \\
\hline
\multirow{3}*{NAOC} & 60/90cm Schmidt telescope & 2354 & 58& 30 & 32\\
  & 40cm double-tube refractor & 2624 & 108 & 55& 37\\
  & No record & 2 & 2 & 2  & - \\
\hline
\multirow{3}*{YNAO} & 1m reflector telescope & 751 & 85 & 16 & 67 \\
  & 60cm reflector telescope & - & 2 & 2  & - \\
  & No record & 2 & 14& -  & - \\
\hline
\multirow{4}*{PMO} & 15cm reflector telescope & 3719 & 340& 284& 36\\
  & 40cm double-tube refractor & 3279 & 111& 66 & 70 \\
  & 60cm reflector telescope & 1114 & 202& 156& - \\
  & No record & 171 & 358& 101& - \\
\hline
\textbf{Total} &  & 15696 & 1883& 1083 & 270 \\
\hline
\end{tabular}}
\label{tab:plate_count}
\end{table}

\begin{figure}
    \includegraphics[width=\linewidth]{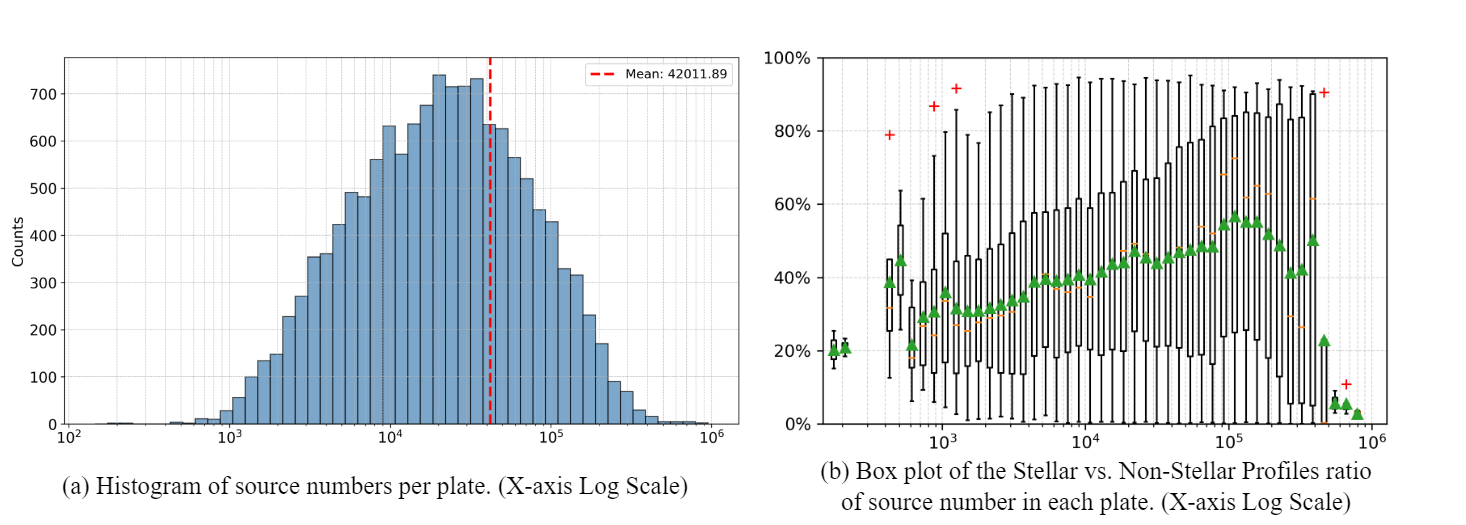}
    \caption{
    Distribution of Plate Count (a) and Stellar Fraction (b) by Source Number Plates. (Boxes represent the interquartile range, central lines indicate medians, red crosses mark outliers, and green triangles denote mean values.)
    % Box plot of the Stellar Fraction of source number (Log Scale) in each plate. (Boxes represent the interquartile range, central lines indicate medians, red crosses mark outliers, and green triangles denote mean values.)
    }
    \label{fig:match_distribution}
\end{figure}

We built our training dataset from \num{15696} astrometrically calibrated plates compiled by \cite{Shang_2024}, originating from five Chinese astronomical institutions: Shanghai Astronomical Observatory (SHAO), National Astronomical Observatories (NAOC), Purple Mountain Observatory (PMO), Yunnan Astronomical Observatory (YNAO), and Qingdao Observatory (QDO) (see Table~\ref{tab:plate_count}).
To obtain reliable labels, we separated stellar from non-stellar sources by cross-matching \texttt{SExtractor} catalogs with \texttt{Gaia DR2} \cite[][]{2018A&A...616A...1G}. The matching radius was initially set at 5$^{\prime\prime}$ and iteratively reduced until it converged to a stable value of the root mean square (rms) of the fitting residuals, or failure to converge constituted a calibration failure. Sources with angular separations smaller than $3\times rms$ from a \texttt{Gaia} counterpart (g band$\leq18$ mag) were treated as matched (positive samples), while all remaining detections (either spurious or galactic sources not in \texttt{Gaia}) were used as negative samples.
% To obtain reliable labels, we separated stellar from non-stellar sources by cross-matching \texttt{SExtractor} detections with \texttt{Gaia DR2} \cite[][]{2018A&A...616A...1G}: sources with angular separations smaller than 2$^{\prime\prime}$ from a \texttt{Gaia} counterpart (g band$\leq18$ mag) were treated as matched (positive samples), while all remaining detections (either spurious or galactic sources not in \texttt{Gaia}) were used as negative samples.
For each source detected by \texttt{SExtractor}, a cutout is created by first forming a bounding box that extends from the source position (X\_IMAGE and Y\_IMAGE parameters) according to its windowed position estimates x and y (XWIN\_IMAGE and YWIN\_IMAGE parameters). This cutout is then extracted and resized to a standardized $64\times64$ pixel patch.
% For each source detected by \texttt{SExtractor}, a cutout was made according to its corresponding bounding box(Sextractor中的什么参数), which was then extracted and resized to a standardized $64\times64$ pixel patch.

We curated a dataset encompassing diverse observational conditions for our foundational model. Figure~\ref{fig:match_distribution} illustrates the distribution of plate count and stellar source fraction by source number across the \num{15696} astrometrically calibrated plates. As shown, the distribution of total sources per plate indicates that the majority of plates contain between approximately $3 \times 10^{3}$ and $10^{5}$ sources, while stellar and non-stellar counts remain relatively balanced in plates with fewer than $3 \times 10^{5}$ total sources. Building a training set with over $10^{4}$ plates, each containing an average of $4\times10^{4}$ sources, obviously exceeds our training hardware capabilities. Additionally, as these plates were obtained from \num{11} telescopes across five different observatories with highly uneven quantities, the complete dataset is not an optimal choice for model training.

Based on these considerations, we implemented a balanced sampling scheme. First, all \num{141} plates from QDO (with the minimum number of plates) were retained to preserve their representation in the dataset. For the remaining four observatories, we randomly selected plates while maintaining an approximately balanced distribution, with the sample comprising n(SHAO) = \num{215}, n(NAOC) = \num{215}, n(YNAO) = \num{214}, and n(PMO) = \num{215}. This results in a final curated training set of \num{1000} plates, with randomly selected \num{2000} positive samples (stellar sources) and \num{2000} negative samples (non-stellar sources) per plate.

% #接下来，我们首先构建一个包含各种情况的训练集来训练我们的基础模型。在图1中，我们给出了15696张已经进行过天体测量定标的底片中恒星和非恒星源数目的分布情况。从图中可看到，大部分底片中的天体数目在？到？之内，其中恒星和非恒星源的数目也基本相当，在。。到范围内。显然，构建一个包含超过10000张底片且每张底片大概40000个源的全样本的训练集超出了我们的硬件能力。另外一方面，这些底片中包含5个不同台站11台望远镜拍摄的底片，且数量并不均衡，这样的全样本训练集也不是一个优化的选择。
% 基于以上考虑，我们训练样本的挑选策略如下。对于15696张底片，QDO的底片一共只有141张，为了数据集的均衡，我们保留了其所有数据。对于SHAO, NAOC, PMO, and YNAO四个台站，我们在它们的所有底片（n=？）中随机挑选了861张。由于这4个台站的底片总数量基本相当，挑选出的样本数目也基本均衡。具体来说，（n(SHAO)= ？？n(NAOC）=?? ..这样，最终的底片数目为？张，5个台站的底片数目基本相当。在这？底片中，正样本（恒星源）多少个，负样本（非恒星样本）多少个。

It is worth noting that the above training set was designed to train a base model to classify stellar and non-stellar sources across all plate types. Nevertheless, this model may perform suboptimally on specific or atypical plates. In such cases, we will adopt a one-to-one fine-tuning approach, where the base model is further adapted to the characteristics of specific plates to enhance classification performance (see Section~\ref{sec:inference_transfer} for details).

% #值得一提的是，以上训练集是为了训练一个基础模型以同时解决所有不同类型底片中的恒星和非恒星源的分类问题。事实上，这样的模型可能对一些特殊的底片仍然是不工作的。在这种情况下，我们将需要采用以一一对应的训练方法来精调基础模型以获取更好的分类能力（详见。。。。）

% this dataset was used to train the base model capable of handling cross‑observatory plate variations. In subsequent stages, specialized fine‑tuning subsets will be constructed by further calibrated plates based on specific observatory characteristics and exposure times, enabling adapted performance on targeted observational regimes. 

\subsection{Astronomical Uncalibrated plates}\label{sec:uncaldata}

\begin{figure}
    \centering
    \includegraphics[width=0.9\linewidth]{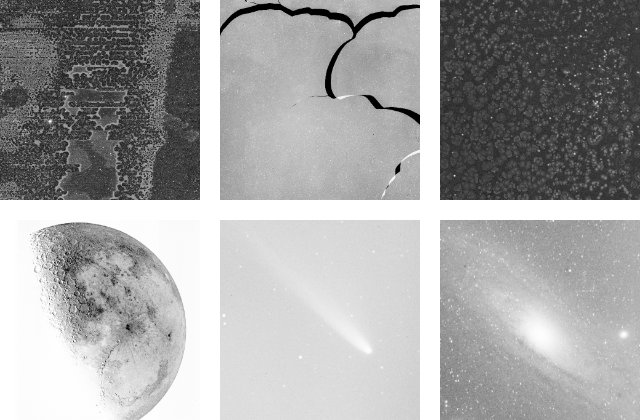}
    \caption{
    Representative defects in photographic plate data, categorized by storage (top row), and include plates with specific astronomical objects (bottom row).
    }
    \label{fig:error_images}
\end{figure}
% #如前文所提，一共还有2530张中国夜天文底片还没有进行天体测量定标。However，并不是每张这样的底片都需要或者能够进行恒星和非恒星的分类的。具体来说，有两类这样的底片，一类是大天体（如Moon，M31，彗星）这些无需定标，这样的底片有多少张（see 图）？另外一类是严重。。。，专业有几张（see 图）.

% 除此之外，237张底片缺少历元，对于这些底片尽管我们可以用分类模型进行分类，但是无法利用Gaia星表进行精确的天体测量定标（see 图2的第几个panel），因此也无法确认分类模型的准确程度，因此也不在本文当中进行讨论。

% In summary，there are ？ 底片需要在本文中进行源的分类。其中，各台站的数目分别给表的column ？中。

As mentioned above, \num{2530} Chinese nighttime astronomical plates remain without astrometric calibration.  Among them, we  exclude two specific plate categories from further process. The first category consists of plates that have been severely degraded during long-term storage or affected by digitization artifacts (see the top row of Fig.~\ref{fig:error_images}). These include: (1) contamination by residual chemicals or fungal growth, (2) physical fractures or partial peeling of the emulsion layer, and (3) pronounced emulsion shrinkage. This group contains \num{443} plates. The second category comprises plates that feature very large astronomical targets (such as the Moon, comets, or M31), for which astrometric calibration is not required (see the bottom row of Fig.~\ref{fig:error_images}); this accounts for an additional \num{204} plates. 

Currently, to ensure scientific reliability, we rely on manual visual inspection to identify and exclude the two excluded categories. Given the relatively limited number of such plates, this approach remains feasible at the present stage. However, for future large-scale processing of archival datasets comprising tens of thousands of plates, manual screening would become impractical. We therefore plan to incorporate automated image quality assessment methods, such as connected-component analysis and global image statistics (e.g., high-intensity pixels of large contiguous regions and image entropy), to automatically flag plates containing oversized targets or severe degradation, thereby significantly reducing manual intervention in future large-scale applications.
%However, not all of these plates need, or are suitable for, high-precision astrometric registration. 
%Although classification models can still be run on these plates, accurate astrometric calibration using the \texttt{Gaia} catalog is impossible without a known epoch (see the final panel in Fig.~\ref{fig:traditional_method}). 

%However, given varying degrees of environmental deterioration, not all of these plates are suitable for high-precision astrometric registration. To systematically assess their usability, we followed Shang et al. (2024) to classify the plates based on emulsion condition, mold extent, and development quality. Under this scheme, Grade 1 plates feature intact emulsions with no mold; Grade 2 plates exhibit minor detachment, mold, or damage covering less than 25\% of the surface area; and Grade 3 plates show similar defects covering up to 50\% of the area. The remaining Grade 4 plates that are damaged and emulsion are deemed unusable. By benchmarking our classification model across the viable plates (Grades 1–3), we aim to demonstrate its robustness in handling various degrees of emulsion damage and mold.

Together, \num{1883} plates remain to be classified in this work. The corresponding plate counts for each observatory are listed in the second column of Table~\ref{tab:plate_count}. Among them, 512 are  well-preserved Grade 1 plates, 728 are  Grade 2 plates exhibiting minor detachment, mold, or damage covering less than 25\% of the surface area, and 643 are Grade 3 plates showing similar defects covering up to 50\% of the area, see \cite{Shang_2024} fore detail.

Finally, it is worth mentioning that there are \num{237} plates missing epoch information. As stellar proper motions are relatively small, we assigned an approximate epoch around 1950 to these plates to substitute for the unknown epoch. 

% Moreover, \num{237} plates are missing epoch information. Although classification models can still be run on these plates, accurate astrometric calibration using the \texttt{Gaia} catalog is impossible without a known epoch (see the final panel in Fig.~\ref{fig:traditional_method}). As a result, the performance of the classification models cannot be rigorously validated for these cases, and they are therefore also omitted from the present analysis. Altogether, \num{1646} plates remain to be classified in this work. The corresponding plate counts for each observatory are listed in the second column of Table~\ref{tab:plate_count}. 

% (这个没有epoch的plate扔掉，我在科学上存疑的，原因是即使50年的历元，自行相比于源匹配的2角秒也是小量，你和于涌老师再讨论一下，比如是不是可以用这些源的平均历元来代替，然后在释放数据里说明一下这些底片没有准确历元)

% With other plates lacking identifiable stellar sources, \num{647} plates were excluded. And proper-motion estimation requires precise epoch information to measure positional shifts over time, \num{237} plates with ambiguous dates were also excluded. The remaining \num{1646} uncalibrated plates were retained for astrometric calibration (Table~\ref{tab:plate_count}). Because these plates originate from multiple telescopes with differing apertures, optical designs, and observing conditions, we grouped them by telescope to mitigate inter-instrument variation. For these uncalibrated plates, the objective is to classify stellar objects directly from the catalogues in the absence of reliable astrometric solutions.

\section{Results}\label{sec:results}

\subsection{Training and evaluation of base model}\label{sec:match_data}

% 我们采用了Adam去优化SwinT模型，并使用循环学习率计划（$3\times10^{-4}$至$10^{-5}$）训练了100多个迭代周期。经过训练的网络是一个通用的分类器，可以从可靠校准的数据中捕获跨观测站的源，以下称为基础模型。该基准测试是在具有以下规格的系统上执行的：CPU–5.2 GHz Intel i9-10900；GPU–英伟达rtx3090（24GB）；RAM–32GB；操作系统-Ubuntu 20.04（64位）。

%he trained network is a general-purpose classifier that captures cross-observatory source from reliably scanned image data, and is hereafter referred to as the \textit{Base\_Model}. 

We train the stellar/non-stellar classification model (Section~\ref{sec:classifier}) using the training dataset introduced in Section~\ref{sec:traindata}, which is referred to as the \textit{Base\_Model} below. The training dataset is separated into training and validation samples with a ratio of 90:10. 
For a binary classification problem with $N$ sources, the $i$-th source has a ground-truth label $y_i \in \{0:non-stellar,1:stellar\}$, and the classifier predicts a stellar probability $p_i \in [0,1]$.
The model is optimized using cross-entropy as the loss function, which is defined as~\ref{equ:cross-entropy}.
\begin{equation}\label{equ:cross-entropy}
    Loss = -\frac{1}{N} \sum_{i=1}^N \left[ y_i \log(p_i) + (1 - y_i) \log(1 - p_i) \right]
\end{equation}

We use the Adam optimizer \citep{2014arXiv1412.6980K} and train the model over \num{100} epochs with a cyclical learning rate schedule ($3\times10^{-4}$ to $10^{-5}$). 
% 这个模型在xx机器上耗时多久训练，训练结束后达到了xx效果。
All experiments were conducted on a system with the following specifications: CPU – 5.2 GHz Intel i9-10900; GPU – NVIDIA RTX 3090 (24GB); RAM – 32 GB; OS – Ubuntu 20.04 (64-bit). Each training epoch takes approximately 894.4 seconds.
For an intuitive evaluation of the model, we classify a source as stellar if it predicts $p_{i}>0.5$, and compute accuracy by comparing these predictions with the ground-truth labels.
To prevent overfitting, we monitor the training process through performance on a validation dataset. After the 72nd epoch, the validation accuracy consistently exceeds 80\%. 
We select the best validation loss performance achieved at epoch 94, which has an accuracy of 85.13\% and a loss of 0.3551, as our model checkpoint and save it for subsequent use.
% 为了防止训练过程中的过拟合，for validation dataset
% 在多少个epoch后，acc表现都高于80；然后再其中挑选验证集效果最好的epoch为我们的base model。

% 然后接下来说模型的效果。
% 需要详细说测试集的效果吗？如果我们是basemodel，是不是只要说一下整体上的效果
% 有这样一个隐含的逻辑：由于我们的分类模型直接关系到的是底片天体测量定标的成果与否，因此利用测试集的评估效果本身并不重要

% 总体而言，\textit{Base\_Model}在所有未校准的板上达到了80.13%的平均精度。同时具有极快的速度，在推理过程中它在NVIDIA RTX 3090 GPU上在\textbf｛7.8秒｝内处理了｛20000｝个源。

\subsection{Inference with base model}\label{sec:unmatch_inference}
% 在｛1646｝个未标准板块中，由于源目录最初缺乏可靠的恒星锚定，与｛Gaia DR2｝的交叉匹配失败。为了最终和gaia匹配成功，首先，为了提高目录质量我们应用\textit{Base\_Model}推理Sextractor的catalog来过滤掉虚假检测。其次，按照章节2.3，我们然后使用\texttt{Astrometry.net}对精确的星表进行了初步的天体测量比对，此步骤成功对于｛1235｝个板（75.03%）。最后，为了导出的世界坐标系（WCS）解，我们将这些以上结果与更新的\texttt{Gaia DR3}\cite[]{2023A&A…674A…1G}进行了交叉匹配。结果共有\textbf{958}个板（原始组的58.20%）确认校准成功。

For these \num{1883} uncalibrated plates in section~\ref{sec:uncaldata}, we follow the astrometric registration flow described in Section~\ref{sec:calibration_workflow}. We first run \texttt{SExtractor} to obtain stellar catalogs and then apply the \textit{Base\_Model} inference on the resized ($64\times64$ pixels) cutout images of \texttt{SExtractor} sources.
The inference was performed at superfast speed, processing \num{20000} sources in \textbf{7.8 seconds} on an NVIDIA RTX 3090 GPU.

%initial cross-matching with \texttt{Gaia DR2} failed because an excessive number of spurious sources in the \texttt{SExtractor} catalogs prevented the identification of reliable stellar anchors. To ultimately achieve successful matching with Gaia, we first applied the 

% After the stellar classification (the stellar sources are selected with $P_{i}>0.5$), we run the astrometrical matching following the procedure in Section~\ref{sec:calibration_workflow}. Specifically, we performed preliminary astrometric alignment using \texttt{Astrometry.net}, followed by a refined cross-match with \texttt{Gaia DR3} catalog \cite[][]{2023A&A...674A...1G}, which required at least 10 matched sources within 1$^{\prime\prime}$.%with a positional error of less than 1 arcsecond.
After stellar classification (the stellar sources are selected with $p_{i}>0.5$), we run the astrometrical matching following the procedure in Section~\ref{sec:calibration_workflow}. Specifically, we performed preliminary astrometric alignment using \texttt{Astrometry.net} to obtain an initial transformation between image pixel coordinates and astronomical coordinates. This transformation was then iteratively refined by cross-matching with the \texttt{Gaia DR3} catalog \cite[][]{2023A&A...674A...1G}. The refinement process continued until the solution converged, or failure to converge signaled an unsuccessful calibration.
The astrometric alignment was successful for \num{1365} plates, and final astrometric matching was achieved for \textbf{1083} plates. We list the number of these final astrometrically registered plates for each telescope in the third column of table~\ref{tab:plate_count}.
% In specfic, we first perform a preliminary astrometric alignment of the stellar catalogs using \texttt{Astrometry.net}, which succeeded for 1,235 plates (75.03\%). After that, we further cross-match ? with the updated \texttt{Gaia DR3} catalog \cite[][]{2023A&A...674A...1G}, requiring at least 10 matched sources with a positional error of less than 1 arcsecond.%什么情况下，最终匹配成功。
% We finally obtain a total number of \textbf{958} plates (58.20\% of the original set) with successful astrometric calibration. Table~\ref{tab:plate_count} presents the match plate counts for each telescope in its third column.

% 具体来说，我们先用A。net做初匹配，然后再与Gaia星表做细匹配。对于我们的xxplate来说，我们1235通过初匹配，958张获得了最终的，即完成了配准。

\subsection{Fine-tune models}\label{sec:inference_transfer}

% 我们评估了基础模型在验证数据集中的跨天文台的泛化能力。验证集中QDO首先被排除在外，因为其有限的板块全被用于训练。表~\ref{tab:acc_count}总结了剩下的四个观测站配准底片的验证性能，主要讨论包含望眼镜记录的底片。
%To benchmark the cross-observatory generalization of \textit{Base\_Model}, 

% 对于这些底片，一方面可能是degreed程度太高，我们的分类模型还不足以对其中的源分类，特别是考虑到不同台站望远镜的psf形状不一样。为了进一步测试基础模型的分类能力，我们利用已经配准但未参与训练的样本构成测试集，将基础模型按台站和望远镜划分的底片，模型评估能力show在表格1中。
% 这个模型中不同台站的表现是不一样的，所以我们要进一步微调。

For the uncalibrated plates, their high distortion level may exceed the classification capability of the \textit{Base\_Model}, particularly considering the varying PSF shapes across telescopes from different observatories. To further evaluate the \textit{Base\_Model} performance, we constructed a test dataset using the remaining calibrated plates from the four observatories (since all plates from QDO were assigned to the training set). 
In the calibration process, stellar sources serve as the primary anchor points; however, matching failures are largely attributed to an excess of contaminating non-stellar sources. Thus, we focus specifically on false positives (FP), instances where non-stellar sources are misclassified as stars. We focus on the metrics of accuracy ($Acc$) and precision ($P$) for stellar sources, as summarized in the base column of Table~\ref{tab:acc_count}, with emphasis on plates that include telescope records \footnote{This distinction is relevant because a subset of the plates in our sample lacks associated telescope metadata ("No record" as referenced in Table~\ref{tab:plate_count}).}. %（这个地方需要一个脚注，读者不明白这句话，因为背景里面不知道有plate没记录望远镜信息）

% ##注意语法时态，尽量用一般现在时
% ##注意astrometric calibration，registeration，matching用法的准确性

% We used the validation dataset comprising the remaining calibrated plates from the four observatories, as all plates from QDO were allocated entirely to the training set. Table~\ref{tab:acc_count} summarizes the validation performance, with focus on plates containing telescope records.

\begin{table}
    \centering
    \caption{Evaluate the classification performance by the Telescopes. The best, second-best, and third-best results are highlighted as \colorbox{first}{first}, \colorbox{second}{second}, and \colorbox{third}{third}.}
    \setlength{\tabcolsep}{1.25mm}{
    \begin{tabular}{cccc|cc}
    \hline
        \multirow{2}*{Observatory} & Telescope & \multicolumn{2}{c|}{Base} & \multicolumn{2}{c}{Fine-tune}  \\
        \cline{3-6}
         & aperture(cm)&  \textbf{$Acc$(\%)} &  \textbf{$P$(\%)} &  \textbf{$Acc$(\%)} &  \textbf{$P$(\%)} \\ \hline
        \multirow{2}*{SHAO} & 40 & \colorbox{first}{84.85} & 26.39 & \colorbox{first}{89.12} & 66.45\\
         & 156 & \colorbox{third}{83.19} & 27.46 & \colorbox{second}{87.95}& 59.79\\ \hline
        \multirow{2}*{NAOC} & 40 & 74.23 & \colorbox{first}{87.98} & 78.81 & \colorbox{first}{92.05} \\
         & 90 & 81.46 & \colorbox{second}{86.75} & 85.78& \colorbox{second}{88.11} \\ \hline
        YNAO & 100 & 76.36 & 14.11 & 84.73& 54.35\\ \hline
        \multirow{3}*{PMO} & 15 & 81.48 & 77.44 & 86.79& 78.58\\
         & 40 & \colorbox{second}{83.25} & \colorbox{third}{84.28} & \colorbox{third}{87.30}& \colorbox{third}{86.69}\\
         & 60 & 75.95 & 63.70 & 80.87 & 71.15 \\ \hline
    \end{tabular}}
\label{tab:acc_count}
\end{table}

NAOC and PMO achieve the highest accuracy and precision, especially on medium-aperture telescope plates, indicating strong feature alignment with the \textit{Base\_Model}. YNAO has the lowest accuracy and precision, likely due to limited plate variability causing model overfitting. These results confirm that performance remains observatory-dependent and may be improved through domain-aware fine-tuning. %Overall, the \textit{Base\_Model} reached an average accuracy of 80.13\% across all test plates.

%The performance of \textit{Base\_Model} varies across different observatories, so we need to proceed with further fine-tuning. For the uncalibrated plates,

To achieve this, we build, for each uncalibrated plate, a corresponding subset used as supervised training data to fine-tune the \textit{Base\_Model}. Concretely, for each uncalibrated plate, we choose a group of five plates observed under comparable physical conditions (similar exposure time, same telescope/observatory) for fine-tuning. It is important to emphasize that, although every plate has an observatory identifier, many lack recorded telescope information and exposure times. Consequently, we assembled the fine-tuning subsets according to the following hierarchical scheme:
\begin{itemize}
\item (1) When both telescope aperture and exposure time were available, we trained a dedicated fine-tuning model for each distinct (aperture, exposure) pair.
\item (2) When exposure times were unavailable, we grouped the subsets solely by telescope aperture and trained an additional fine-tuning model for each telescope.
\item (3) For plates missing both telescope and exposure metadata, we trained a single model for each observatory to capture station-level, site-specific effects.
\end{itemize}

While this metadata-driven hierarchical scheme provides robust methods. But, for plates lacking comprehensive observation records (e.g., the "No record" cases in Table \ref{tab:plate_count}), relying solely on metadata is limited. One possible approach for future automated processing is to utilize intrinsic image features, such as the shape of the PSF and the background noise distribution, to guide model selection. By evaluating the similarity of these visual features, the system could automatically match an uncalibrated plate with the most suitable fine-tuned model, thereby reducing the reliance on the plate's metadata record.

% We conducted \num{150} transfer learning according to the protocol described above. 
For each fine‑tuning model, we train it for $30$ epochs with a learning rate of $1\times10^{-5}$ to produce a condition‑specific \textit{Fine‑tuned\_Model}, and we select the checkpoint with the lowest cross-entropy loss as the final result.

\subsection{Inference with Fine-Tune Model}\label{sec:unmatch_inference_transfer}

We first evaluated the performance of \textit{Fine‑tuned\_Model} on test plates, as summarized in the Fine‑tune column of Table~\ref{tab:acc_count}.
The results confirm that \textit{Fine‑tuned\_Model} improves classification performance over the \textit{Base\_Model}, thereby demonstrating its effectiveness in adapting to specific observational conditions.

Subsequently, we applied each condition-specific fine-tuned model to classify stellar objects among the detected sources of the \num{800} unregistered plates. Astrometric alignment was then performed with \texttt{Astrometry.net}, successfully processing \num{320} plates. Finally, cross-matching with \texttt{Gaia DR3} confirmed robust WCS solutions for \textbf{270} plates.

The number of plates obtained with new astrometric registration and fine-tuned classification methods for each telescope is also summarized in the fourth column of Table~\ref{tab:plate_count}. Combined with the \textbf{1083} plates previously calibrated by the base model, a total of \textbf{1353} plates have now been successfully registered. %The details of these plates, including their coordinates and identifiers, are provided in Table~\ref{tab:plate_ok}.

\section{Discussion}\label{sec:discussion}

\begin{table}[h]
\centering
\caption{Astrometric registration performance across different plate quality grades.}
\begin{tabular}{lccc}
\hline
Plate Quality Grade & Total Plates & Successfully Registered & Success Rate \\
\hline\noalign{\smallskip}
Grade 1 (Intact) & 512 & 420 & 82.0\% \\
Grade 2 (Minor defects) & 728 & 494 & 67.9\% \\
Grade 3 (Moderate defects) & 643 & 439 & 68.3\% \\
\hline
Total & \textbf{1883} & \textbf{1353} & \textbf{71.9\%} \\
\hline
\end{tabular}\label{tab:quality_grades}
\end{table}

Table~\ref{tab:quality_grades} shows the final astrometric registration success rates across viable plate grades, quantifying our pipeline's ability to recognize degraded data. Crucially, these plates failed in the traditional methods utilized by \cite{Shang_2024}. And the fine-tuning model was applied exclusively to plates unresolved by the base model, the intrinsic plate sample difficulty varies significantly. Thus, rather than comparing model variants, Table~\ref{tab:quality_grades} highlights the overall robustness of classification pipelines to environmental deterioration.

Our framework successfully calibrated a substantial fraction of these previously discarded traditional methods. For Grade 1 plates, the pipeline achieved an 82.0\% success rate. More importantly, for plates suffering from visible environmental degradation, the model demonstrated exceptional resilience, yielding success rates of 67.9\% and 68.3\% for Grades 2 and 3, respectively. This performance plateau is scientifically significant: it indicates that once the network learns to bypass localized artifacts, it maintains robust feature extraction even as the spatial coverage of defects doubles (from $<$25\% to $<$50\%).

\begin{figure}
    \includegraphics[width=\linewidth]{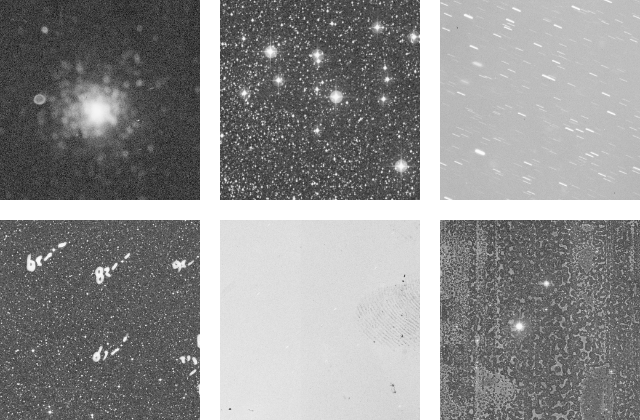}
    \caption{
    Representative Examples of Unclassifiable Stars from Unmatched Plates.
    }
    \label{fig:unmatch_images}
\end{figure}

According to the results of Section~\ref{sec:results}, there are still \num{630} Chinese nighttime single-exposure historical plates remaining to complete the astrometric registration. To understand the reason for the failures, we conducted a detailed inspection of these unregistered plates.
Figure~\ref{fig:unmatch_images} illustrates two types of categories. The top row illustrates issues arising from the observational process, including defocused exposures that result in diffuse stellar profiles, crowded star fields, and streaked or elongated trails caused by mechanical drift during extended exposures. These intrinsic optical defects account for the 18.0\% failure rate observed among the Grade 1 plates. The bottom row displays degradation introduced during plate storage and digitization, such as handwritten annotations, fingerprint smudges, and surface contamination that produces banding or blotching artifacts. These artifacts, which predominantly plague the failed Grade 2 and Grade 3 plates, frequently obliterate the source signals entirely. Taken together, these effects alter the photometric and morphological characteristics of stellar objects to such an extent that reliable classification is not possible, even when using the modern deep learning classifier employed in this study.

%, explaining many unmatched cases.

%Such examples highlight the inherent complexity of historical photographic data, where both observational imperfections and digitization artifacts introduce non-astrophysical structures. Irregular backgrounds, flux saturation, and blended or elongated sources often mislead detection algorithms. These factors reveal the delicate balance between astronomical signals and artifacts in early photographic material, underscoring the need for robust data preprocessing and model adaptation in downstream calibration tasks. 

%Overall, this two-stage pipeline comprises (1) a general calibration via the \textit{Base\_Model} and (2) condition-specific refinement calibration via the \textit{Fine-tuned\_Model}, yielding accurate astrometric solutions for a total of \textbf{1100} additional plates. The resulting calibrated plates, including scanned images, WCS parameters, and stellar source catalogs, will be publicly released to support future studies of historical sky surveys and long-term variability analyses.

\section{Summary}\label{sec:summary}

In this study, we present a deep learning assisted framework to improve the astrometric registration of degraded Chinese historical nighttime astronomical plates. Specifically, a Swin Transformer–based classifier is employed to identify and filter spurious stellar sources, and those astrometrically calibrated plates of \citet{Shang_2024} are used to build the training sample. Besides training a base model based on a mixture sample of over 4 million stellar sources and non-stellar sources from the plates of different telescopes at different sites, we also fine-tune the model with small subsets of samples grouped by metadata (e.g., exposure time, telescope configuration) to adapt the models to different observing conditions. This combined approach provides final astrometric solutions for \textbf{1353}; these plates had previously failed with traditional pipelines, demonstrating that deep learning serves as an excellent classifier for this complex binary classification problem. These data will be publicly released via the NADC and can also be obtained by directly contacting the corresponding authors.

In the future, for similar tasks on the scanned images of astrometric plates, we could incorporate object detection models in deep learning (e.g., YOLO) to locate reliable sources directly, moving toward an end‑to‑end system that eliminates reliance on \texttt{SExtractor} and classifiers. %Further integration of cross‑plate photometric normalization would also help stabilize photometric zero‑points across the archive (##这句话啥意思##. 
Together, these improvements would support more systematic use of historical plates for astrometric and photometric studies.

\section*{Acknowledgements}

This work is supported by the National Natural Science Foundation of China (Grant No. 12473070) and the International Partnership Program of the Chinese Academy of Sciences (Grant No. 018GJHZ2023110GC).

\label{lastpage}
\end{CJK*}
\end{document}